# A thermodynamic approach to the fragility of glass-forming polymers

D. Cangialosi,[1] A. Alegría,[2] J. Colmenero[1,2]

[1]*Fundacion Donostia International Physics Center, Paseo Manuel de Lardizabal 4, 20018 San Sebastián, Spain*

[2]*Departamento de Física de Materiales, Universidad del País Vasco (UPV/EHU) y Unidad de Física de Materiales Centro Mixto (CSIC-UPV/EHU), Facultad de Química, Apartado 1072, 20080 San Sebastián, Spain*

ABSTRACT

We have connected the dynamic fragility, namely the rapidity of the relaxation time increase upon temperature reduction, to the excess entropy and heat capacity of a large number of glass-forming polymers. The connection was obtained in a natural way from the Adam-Gibbs equation, relating the structural relaxation time to the configurational entropy. We find a clear correlation for a group of polymers. For another group of polymers, for which this correlation does not work, we emphasise the role of relaxation processes unrelated to the $\alpha$ process in affecting macroscopic thermodynamic properties. Once an essentially temperature independent contribution of these processes is removed from the total excess entropy, the correlation between dynamic fragility and thermodynamic properties is re-established.

A liquid cooled down below its melting temperature undergoes an increase of the relaxation time (and the viscosity), which eventually leads to the glass transition. Depending on the rapidity of the relaxation time increase supercooled liquids are classified into strong and fragile,[1] respectively displaying mild and rapid variation of the relaxation time with temperature. The connection between the rapidity of variation of the relaxation time of the $\alpha$ process, the so-called dynamic fragility, with other properties of the glass-former, such as thermodynamic,[2,3] vibrational[4] and mechanical properties,[5] has been the subject of intense research in recent years. Following one of these approaches, Angell and co-workers have argued that the dynamic fragility, must be connected to the rapidity of configurational entropy variation, as a natural consequence of the Adam and Gibbs (AG) conjecture that the relaxation time of the $\alpha$ process is correlated to the configurational entropy of the glass-former.[6] A correlation was found for a large number of low molecular weight glass-formers between dynamic and thermodynamic fragility.[2,3] In these works, the dynamic fragility was defined as the ratio between the temperature where the relaxation time is halfway between its value at the glass transition temperature ($T_g$) (~$10^2$ s) and the value at infinite temperature (~$10^{-14}$ s), and the thermodynamic fragility as the ratio of the temperature where the excess entropy is ½ or ¾ of the entropy of fusion and the melting temperature. Despite the general success to connect the so-defined dynamic and thermodynamic fragilities for low molecular weight glass formers, Ngai and co-workers[7,8] and McKenna and co-workers[9,10] have noticed that such a correlation does not hold when the slope of the relaxation time at $T_g$, taken as representative of the dynamic fragility, and the ratio of the liquid by the glass heat capacity at $T_g$, empirically used as a measure of the thermodynamic fragility, are compared for polymeric glass-formers. This conclusion has raised serious criticism about the validity of thermodynamic approaches to the glass transition. More recently, Wang and Angell[11] have used an empirical expression connecting the dynamic fragility to the ratio of the excess specific heat and the enthalpy of fusion, which has been rationalised in the framework of the random first order transition theory of Wolynes and co-workers.[12]

In this work, we revisit the connection between the rapidity of the relaxation time variation and thermodynamic properties starting from AG equation:[6]

$$\tau = \tau_0 \exp\left(\frac{\Delta\mu s_c^*}{k_B T S_c}\right) = \tau_0 \exp\left(\frac{C}{T S_c}\right) \qquad (1)$$

where $k_B$ is the Boltzmann constant, $T$ is the temperature, $S_c$ is the configurational entropy, $\Delta\mu$ is the energy barrier per particle over which a cooperative rearranging group must pass, $s_c^*$ is the configurational entropy associated to such a rearrangement, and $\tau_0$ is the relaxation time at infinite temperature. Both $s_c^*$ and $\Delta\mu$ are assumed to be independent of temperature.

Due to impossibility of accessing experimentally configurational entropy data for glass-forming liquids, the liquid entropy in excess to the corresponding crystal has been used throughout the whole study. Although the two quantities are never the same due to the contribution to the excess entropy of excess vibrations, several recent simulation and experimental studies indicate that proportionality between configurational and excess entropy exists, at least in the relevant range of application of the AG equation.[13-15] A similar proportionality was very recently found also between relaxational and vibrational compressibility.[16] Under this assumption, equation (1) can be rewritten as:

$$\tau = \tau_0 \exp\left(\frac{C'}{TS_{ex}}\right) \qquad (2)$$

where $S_{ex}$ is the excess entropy and $C'$ is proportional to $C$, defined in equation (1). The dynamic fragility can be defined according to Speedy:[17]

$$m_A = \left.\frac{d\left[\frac{\ln(\tau(T)/\tau_0)}{\ln(\tau(T_g)/\tau_0)}\right]}{d\left(\frac{T_g}{T}\right)}\right|_{T=T_g} \qquad (3)$$

Insertion of the AG equation in equation (3) provides a connection between the dynamic fragility and thermodynamic properties:

$$m_A = 1 - S_{ex}(T_g)^{-1}\left[(dS_{ex}(T)/dT)\left(dT\Big/d\left(\frac{T_g}{T}\right)\right)\right]_{T=T_g} \qquad (4)$$

As is possible to see in equation (4), the definition of dynamic fragility provided by equation (3) has the advantage of being insensitive to the dependence of $C$ (or $C'$) on the material, when combined with the AG relation. Rearranging equation (4), one obtains a correlation between the dynamic fragility and the excess heat capacity and entropy:

$$m_A = 1 + \frac{\Delta c_p(T_g)}{S_{ex}(T_g)} \qquad (5)$$

Here $\Delta c_p$ is the excess heat capacity. A similar expression was recently obtained also by other authors[18-20] and successfully verified for a large number of low molecular weight glass formers.[18] Equation (5) is also qualitatively (though not quantitatively) analogous to that proposed by Wang and Angell,[11] and Stevenson and Wolynes.[12]

In this work, we have tested the validity of equation (5) for a large number of polymeric glass-formers. We have found that a correlation exists between dynamics fragility and thermodynamic properties for some polymers, whereas this correlation fails for others. We have attributed the absence of correlation for these polymers to the non-negligible contribution to the excess entropy of all sorts of intra-molecular motions, not related to the $\alpha$ process. This contribution roughly equals the residual excess entropy at the Vogel temperature ($T_0$), namely the temperature where the relaxation time of the $\alpha$ process would diverge. The polymers tested are summarised in table 1. Structural relaxation data were all taken from ref. 21 and thermodynamic data from the ATHAS database and the extensive work cited therein.[22]

In figure 1, the test of equation (5) is displayed for all investigated polymers. As can be seen, a correlation between dynamic fragility and thermodynamic properties, according to the prediction of equation (5) is found for a certain group of polymers (filled circles). On the other hand, the values of the dynamic fragility obtained by equation (5) are too small when compared to the experimental dynamic fragility for another group of polymers (open triangles). In addition, we notice that polymers, for which equation (5) does not hold, have a residual excess entropy at $T_0$, as highlighted by ourselves in a recent paper on the applicability of the AG equation.[21] This is reflected by the fact that, for these polymers, $T_0$ is significantly higher than the Kauzmann temperature ($T_K$), namely the temperature where the excess entropy would extrapolate to zero (see table 1).

These results could in principle be interpreted as a failure of the AG theory connecting dynamics and thermodynamics of glass-forming systems and/or the inability of excess entropy to replace the configurational entropy. However, we notice that correlations recently presented for low molecular weight glass-formers between the dynamic fragility and either vibrational properties[4] or macroscopic properties like the shear or bulk moduli and the Poisson's ratio,[5] clearly fail for some glass-forming polymers.[16,23,24] The main drawback of all these approaches is that they attempt to find a correlation between the dynamic fragility, which is exclusively related to the $\alpha$ process, with properties representative of the global structure of the glass-former. This is a quite reasonable assumption for low molecular weight glass-former and for some polymers with simple monomeric structure. However, this may not be adequate for polymers possessing some kind of intra-molecular degrees of freedom, which may manifest as sub-$T_g$ relaxation processes unrelated to the $\alpha$ process.[25] In this case, if dynamic fragility is connected to thermodynamic properties through equation (4), only the contribution of the $\alpha$ process must be taken into account. Therefore, we can follow a similar approach to that used in ref. 21, where, after removing the contribution to the excess entropy unrelated to the $\alpha$ process, the AG equation could be fitted in wide temperature range above $T_g$, and rewrite equation (5) as:

$$m_A = 1 + \frac{\Delta c_p(T_g)}{S_{ex}(T_g) - S_{ex,\beta}} = 1 + \frac{\Delta c_p(T_g)}{S_{ex,\alpha}(T_g)} \tag{6}$$

Here $S_{ex,\beta}$ has been referred for simplicity as $\beta$ though, as just mentioned, it can in principle be related to all processes unrelated to the $\alpha$ process contributing significantly to the total excess entropy. Moreover, we have assumed this contribution to the excess entropy to be temperature independent (no effect on the heat capacity). This assumption turns out to be reasonable because the values that one needs to subtract from the total excess entropy at $T_g$ to re-establish the positive correlation predicted by equation (5), resemble the excess entropy at the Vogel temperature, namely $S_{ex,\beta} \approx S_{ex}(T_0)$. This is shown in figure 1, where a correlation between dynamic fragility and thermodynamic properties is actually re-established once the excess entropy at $T_0$ is removed from the total excess entropy at $T_g$. The temperature independence of the non-$\alpha$ related excess entropy is in agreement with the interpretation that this contribution derives from intra-molecular motions along the polymer chain. In this case, the number of possible configurations available would be essentially temperature independent. An additional support to this temperature independence is supplied by the absence of sub-$T_g$ specific heat jumps for those polymers for which equation (6) applies. The scenario arising from the application of equation (6) is displayed in figure 2, where the total excess entropy together with the contribution from the $\alpha$ process and the other processes is drawn schematically as a function of temperature. As can be observed, according to this scenario the excess entropy from the $\alpha$ process would vanish at $T_0$, where significant excess entropy is still present.

Apart from the validation of the connection between relaxation time and excess entropy, the presence of relaxation processes unrelated to the $\alpha$ process contributing to the global properties of some polymers offers a plausible explanation to the inability of other approaches to maintain a clear correlation between dynamic fragility, and vibrational[4] or mechanical properties.[5] However, it is noteworthy that, the latter approaches do not allow distinguishing between contributions to global properties coming from different relaxation processes, whereas in this study the extrapolated residual excess entropy at $T_0$ offers an elegant route to quantify the non-$\alpha$ process contribution to global properties.

To summarise, we can conclude that the mere comparison between the dynamic fragility, a property exclusively related to the $\alpha$ process, and properties representative of the overall structure of the glass former, such as the excess entropy (and its temperature variation), the compressibility[5,16,23] and the vibrational properties,[4] may result in some inconsistencies. These may arise from the presence of contributions to the overall system properties that are not related to the $\alpha$ process (for which the dynamic fragility is defined), such as intra-molecular motions in glass-forming polymers.

REFERENCES

1. C. A. Angell, In Relaxations of Complex Systems, edited by K. L. Ngai, G. B. Wright (National Technical Information Service, U.S. Department of Commerce, Springfield, VA 1985).

2. K. Ito, C. T. Moynihan, and C. A. Angell, Nature **398**, 492 (1999).

3. L. M. Martinez and C. A. Angell, Nature **410**, 663 (2001).

4. T. Scopigno, G. Rocco, F. Sette, and G. Monaco, Science **302**, 849 (2003).

5. V. N. Novikov and A. V. Sokolov, Nature **431**, 961 (2004); J. C. Dyre, Nature Mat. **3**, 749 (2004).

6. G. Adam and J. H. Gibbs, J. Chem. Phys. **43**, 139 (1965).

7. C. M. Roland, P. G. Santangelo, and K. L. Ngai, J. Chem. Phys. **111**, 5593 (1999).

8. K. L. Ngai and O. Yamamuro, J. Chem. Phys. **111**, 10403 (1999).

9. D. Huang and G. B. McKenna, J. Chem. Phys. **114**, 5621 (2001).

10. D. Huang, D. M. Colucci, and G. B. McKenna, J. Chem. Phys. **116**, 3925 (2002).

11. L. M. Wang and C. A. Angell, J. Chem. Phys. **118**, 10353 (2003).

12. J. D. Stevenson and P. G. Wolynes, J. Phys. Chem. B **109**, 15093 (2005).

13. C. A. Angell and S. Borick, J. Non-Cryst. Solids **307**, 393 (2002) and references therein.

14. D. Prevosto, M. Lucchesi, S. Capaccioli, R. Casalini, P. A. Rolla, Phys. Rev. B **67**, 174202 (2003); D. Prevosto, S. Capaccioli, M. Lucchesi, D. Leporini, and P. Rolla, J. Phys.: Condens. Matter **16**, 6597 (2004).

15. S. Corezzi, L. Comez, and D. Fioretto, Eur. Phys. J. E **14**, 143 (2004).

16. U. Buchenau and A. Winschewski, Phys. Rev. B **70**, 092201 (2004).

[17] R. J. Speedy, J. Phys. Chem. B **103**, 4060 (1999).

[18] U. Mohanty, N. Craig, J. T. Fourkas, J. Chem. Phys. **114**, 10577 (2001).

[19] R. Casalini, M. Paluch, T. Psurek, C. M. Roland, J. Mol. Liq. **111**, 53 (2004).

[20] G. Ruocco, F. Sciortino, F. Zamponi, C. De Michele, T. Scopigno, J. Chem. Phys. **120**, 10666 (2004).

[21] D. Cangialosi, A. Alegría, and J. Colmenero, Europhys. Lett. **70**, 614 (2005) and references therein.

[22] B. Wünderlich, ATHAS database, http://web.utk.edu/~athas/ and references therein.

[23] V. N. Novikov and A. V. Sokolov, Phys. Rev. E **71**, 061501 (2005); U. Buchenau, 5th International Discussion on Relaxation of Complex Systems, Lille 7-13 July 2005.

[24] In this work, PVC belongs to those polymers for which a correlation between dynamic fragility and excess entropy exists without invoking the contribution of processes unrelated to the $\alpha$ process. This is in apparent contradiction with the absence of correlation between fragility and the ratio between vibrational and relaxational compressibility found in ref. 16 for PVC. However, in the present case, PVC dynamic fragility is evaluated in the hypothesis of a heterogeneous structure of PVC as proposed by: A. Arbe, A. Moral, A. Alegria, J. Colmenero, W. Pyckhout-Hintzen, D. Richter, B. Farago, and B. Frick, J. Chem. Phys. **117**, 1336 (2002).

[25] K. L. Ngai and M. Paluch, J. Chem. Phys. **120**, 857 (2004).

| Polymer | $T_g$ (K) | $T_k$ (K) | $T_0$ (K) | $S_{ex}(T_0)$ (J/molK) |
|---|---|---|---|---|
| i-PP | 270 | 174.5±3.2 | 174±7 | 0±1.1 |
| LDPE | 237 | 148.9±1.2 | 160.6±3.5 | 1±0.3 |
| cis-PI | 200 | 166.5±2.3 | 162±13 | -0.9±2.6 |
| PVC | 354 | 308±13 | 317±3.9 | 0.38±0.2 |
| PDMS | 146 | 130.3±0.5 | 130±0.2 | 0±0.05 |
| PS | 373 | 278±5.6 | 324±2.5 | 6.7±0.3 |
| PEN | 390 | 325.8±2.2 | 358±1.6 | 9.1±0.4 |
| PET | 342 | 269±13 | 308±1.1 | 14.0±0.3 |
| a-PMMA | 378 | 263.3±3.5 | 334±15 | 10.7±1.8 |
| PC | 420 | 292±6.5 | 373±2.6 | 21.7±1.4 |
| PEEK | 419 | 324±10 | 398.4±3 | 22.5±0.7 |
| PA 6,6 | 323 | 196±8.6 | 290±16 | 64.0±6.8 |

Table 1

CAPTIONS FOR FIGURES

FIGURE 1: Comparison between the experimental dynamic fragility and that obtained from the application of the AG equation. Filled circles correspond to polymers that verify equation (5) and open triangles to polymers for which equation (5) fails. Filled triangles correspond to polymers for which equation (6) was applied, i.e. after subtraction from the total excess entropy of the contribution unrelated to the $\alpha$ process.

FIGURE 2: Schematic representation of the scenario arising from the application of equation (6) for those polymers for which the total excess entropy is not representative for the excess entropy related to the $\alpha$ process.

FIGURES

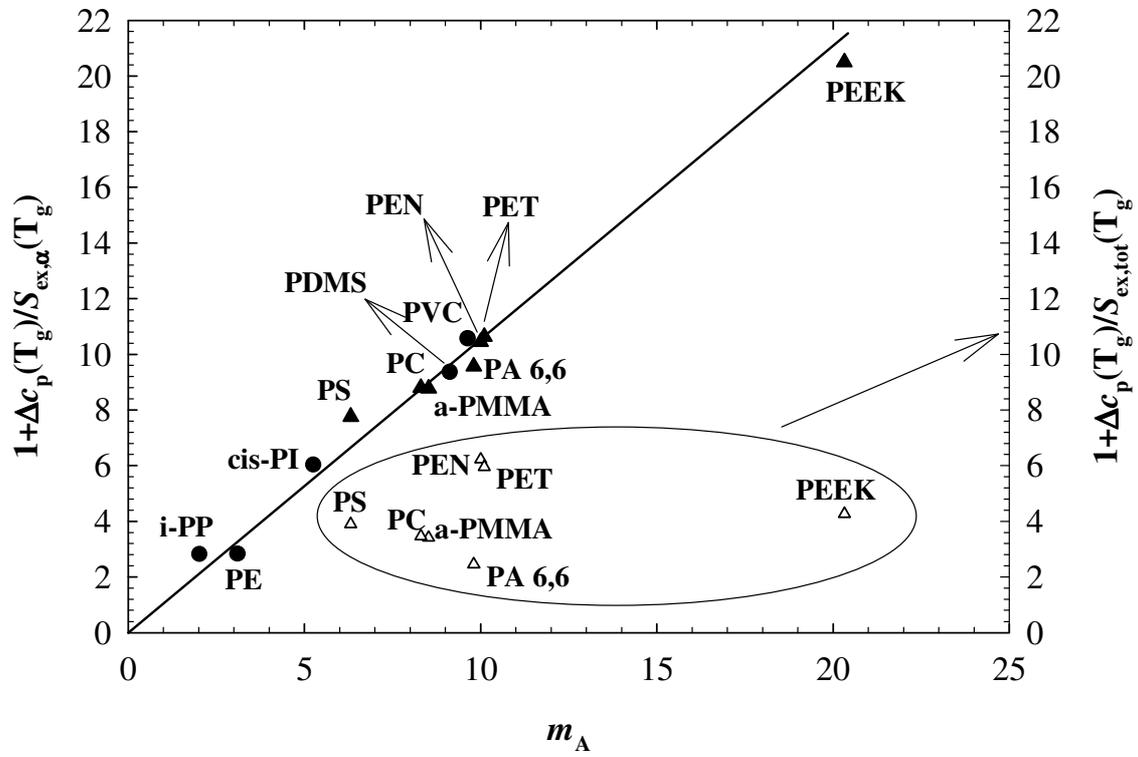

Figure 1

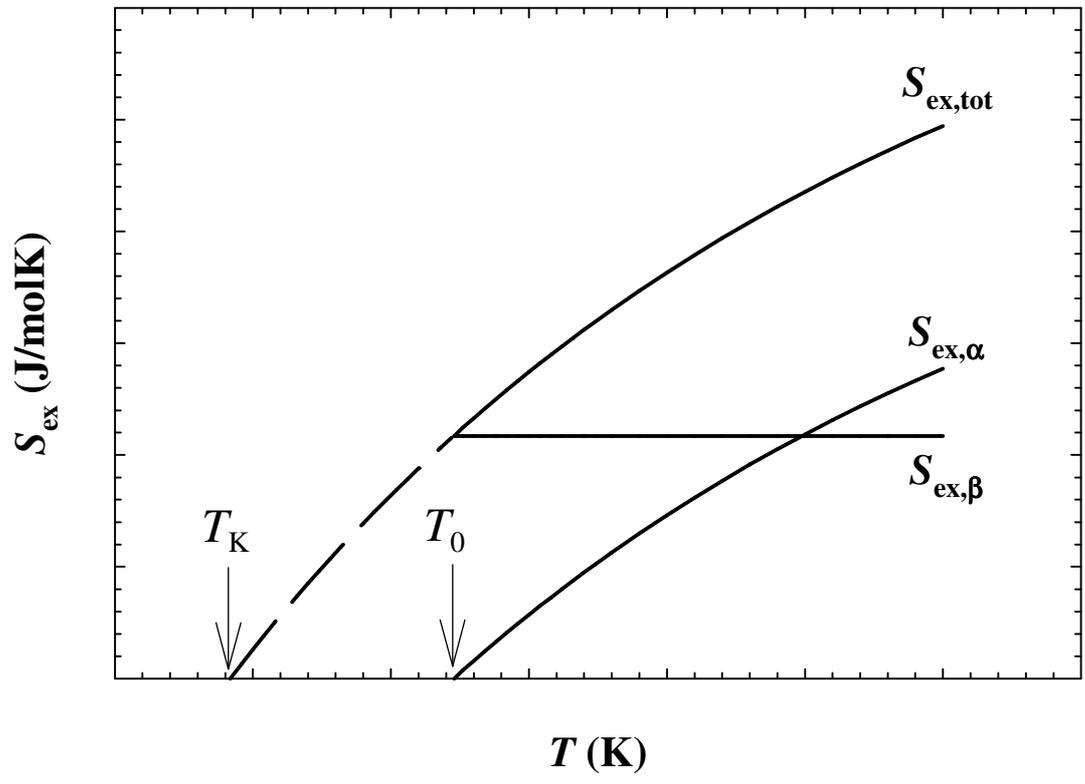

Fig. 2